\documentclass{article}

\usepackage{graphicx}
\usepackage{natbib}
\usepackage{amsmath}
\usepackage{cases}   
\usepackage{indentfirst}
\usepackage{amsfonts}
\usepackage{amssymb}
\usepackage{makeidx}
\usepackage{latexsym}
\usepackage{bbm}
\usepackage{mathrsfs}
\usepackage{varioref}
\usepackage[applemac]{inputenc}

\title{Study on Coulomb explosions of ion mixtures}
\usepackage{authblk}
\author[1]{E. Boella}
\author[2]{B. Peiretti Paradisi}
\author[3]{A. D'Angola}
\author[2]{G. Coppa}
\author[1]{L. O. Silva}
\affil[1]{GoLP/Instituto de Plasmas e Fusao Nuclear, Instituto Superior Tecnico, Universidade de Lisboa, Avenida Rovisco Pais, 1, Lisbon, Portugal}
\affil[2]{Dipartimento Energia, Politecnico di Torino, corso Duca degli Abruzzi, 24, Torino, Italy}
\affil[3]{Scuola di Ingegneria, Universita della Basilicata, via dell'Ateneo Lucano, 10, Potenza, Italy}

\date{}
\begin{document}
\maketitle
\begin{abstract}
The paper presents a theoretical work on the dynamics of Coulomb explosion for spherical nanoplasmas composed by two different ion species. Particular attention has been dedicated to study the energy spectra of the ions with the larger charge-to-mass ratio. The connection between the formation of shock shells and the energy spread of the ions has been the object of a detailed analysis, showing that under particular conditions the width of the asymptotic energy spectrum tends to become very narrow, which leads to a multi-valued ion phase-space. The conditions to generate a quasi mono-energetic ion spectrum have been rigorously demonstrated and verified by numerical simulations, using a technique that, exploiting the spherical symmetry of the problem, allows one to obtain very accurate and precise results. 
\end{abstract}


\section{Introduction}
The interaction between sufficiently intense laser pulses and spherical nanoclusters can lead to the complete expulsion of all the electrons from the cluster, as discussed in \citep{Ditmire-PRA-1996, Murakami-PoP-2009, Coppa_simple}, and to a consequently rapid explosion of the left-behind ion cloud. The process is a well known mechanism of ion acceleration \citep{Ditmire-NAT-1997} with many potential applications, ranging from fusion research \citep{Ditmire-NAT-1999} to biomolecular imaging \citep{Neutze-NAT-2000}. While the explosion of a single species ion plasma leads to a broad energy spectrum \citep{Last-PRA-2001, Krainov-PRA-2001}, it has been verified that in the presence of ion mixtures conditions exist producing a quasi-monoenergetic distribution of the species having the larger charge-to-mass ratio \citep{Last-PRL-2001, Last-PRA-2005, Hohenberger-PRL-2005, Murakami-PoP-2008, Andreev-PoP-2010, Last-EPJD-2009, Last-PoP-2010}. Numerical simulations also pointed out a connection between the monochromaticity of the spectrum and the occurrence of overtakings among faster ions, leading to the formation of shock shells \citep{Popov-PoP-2010, Li-Liu-2007, Last-EPJD-2010}.

The present work investigates in detail and from a theoretical point of view the energy spectrum of charged particles accelerated during the Coulomb explosion of ion mixtures. In particular, a mixture composed by two types of ions uniformly distributed is considered. The ions of the two species have masses $m_s$, $m_f$ and electric charges $q_s$, $q_f$, with 
\begin{equation}
\gamma=\dfrac{q_s/m_s}{q_f/m_f}<1\textrm{ .}
\end{equation}
The ions, which are initially at rest, start moving under the influence of the repulsive electrostatic forces, and the species with a larger charge-to-mass ratio moves faster. In this way, two concentric spherical regions, $S_f$ and $S_s$, are created having radius $R_f(t) \text{ and } R_s(t)$ respectively, with $R_s \leqslant R_f$. The sphere $S_s$ contains fast and slow particles and $R_s$ represents the frontline of the slow ions; instead, the spherical shell in $S_f$ outside $S_s$ contains only fast particles.
The ion dynamics is different in these two regions, and it is studied in detail in Sect. 2 (inner region) and in Sect. 3 (outer region). The dynamics of the explosion depends strongly on the presence of shock shells. It is proved rigorously that the appearance of shocks depends only on the ratio between the total charge of the fast ions and the one of the slow ions, and not on their masses (Sect. 4). In Sect. 5, by using a Hamiltonian approach the properties of the energy spectra are studied analytically.
In particular, a rigorous proof is provided for the existence of a threshold for the generation of shock shells (previously this proof was known only for the limit situation of  $m_s \rightarrow+\infty$ \citep{Li-Liu-2007}), so confirming the results of numerical simulations \citep{Murakami-PoP-2009}.
 Finally, some conclusions are reported in Sect. 6.

Whenever possible, the properties of the explosions have been deduced analytically, otherwise the study has been performed numerically with the so called ``shell method'' \citep{APS-Poster, Dangola-PoP-2014}, which provides extremely precise results by exploiting the spherical symmetry of the problem.
 
\section{Ion expansion in the inner zone}\label{sec:Ion expansion in the inner zone}
Initially, all the ions are uniformly distributed inside a sphere of radius $R$. As the charge density is constant, the radial electric field is linear. Therefore, there will be a uniform expansion for both slow and fast ions, and consequently the charge density will remain constant inside the sphere $S_s$, producing again a linear behaviour of the electric field. From this consideration, one can infer that the electric field inside $S_s$ remains a linear function of the radius $r$ at any time:
\begin{equation}
E(r,t)=A(t)r  \quad \text{for} \quad r \leqslant R_s(t). \label{linear_efield}
\end{equation}
Under this assumption, the equations of motion for the fast and the slow ions inside $S_s$ can be written in a simple way. 
The uniform expansion can be described by introducing two functions, $\xi_s(t)$ and $\xi_f(t)$, such that  slow ions and fast ions  initially at $r=r_0$ are moved at time $t$ to $r=r_0 \xi_s(t)$ and $r=r_0 \xi_f(t)$, respectively. Moreover, the initial densities of the two species, $n_{s,0}$ and $n_{f,0}$, evolve in time as 

\begin{equation}
n_s(t)=\dfrac{n_{s,0}}{\xi_s^3(t)},\: n_f(t)=\dfrac{n_{f,0}}{\xi_f^3(t)}.
\end{equation}
Using Gauss's law, the electric field inside the sphere $S_s$ can be written as: 
\begin{equation}
E(r,t)=\frac{4\pi}{3}\left(\frac{q_fn_{f,0}}{\xi_f^3(t)}+\frac{q_sn_{s,0}}{\xi_s^3(t)}\right)r,
\label{Efield}
\end{equation}
which has the correct dependence on $r$, as in Eq. \eqref{linear_efield}.
After introducing expression \eqref{Efield} into the equations of the motion for fast and slow ions, by writing the acceleration of the ions of the two species as $r_0 d^2\xi_f/dt^2$ and $r_0 d^2\xi_s/dt^2$, one finally obtains
\begin{eqnarray}
\left \{
\begin{matrix}
\dfrac{ \textrm{d}^2 \xi_f}{\textrm{d} t^2} &=& \dfrac{4 \pi}{3} \dfrac{q_f}{m_f} \left(\dfrac{q_fn_{f,0}}{\xi_f^3}+\dfrac{q_sn_{s,0}}{\xi_s^3}\right) \xi_f \textrm{ ,} \\
& & \\
\dfrac{ \textrm{d}^2 \xi_s}{\textrm{d} t^2} &=& \dfrac{4 \pi}{3} \dfrac{q_s}{m_s} \left(\dfrac{q_fn_{f,0}}{\xi_f^3}+\dfrac{q_sn_{s,0}}{\xi_s^3}\right) \xi_s\textrm{ ,}
\end{matrix}
\right .  \label{system_to_solve}
\end{eqnarray}
in which  $r_0$ does not appear.

Equations \eqref{system_to_solve} can be rewritten in a more compact way, as 
\begin{eqnarray}
\left \{
\begin{matrix}
\dfrac{ \textrm{d}^2 \xi_f}{\textrm{d} t^2} &=& \nu^2 \left(\dfrac{\alpha}{\xi_f^3}+\dfrac{\beta (1-\alpha)}{\xi_s^3}\right) \xi_f, \quad&\xi_f(0)=1,&\quad &\dfrac{\textrm{d}\xi_f}{\textrm{d}t}(0)=0& \\
& & \\
\dfrac{ \textrm{d}^2 \xi_s}{\textrm{d} t^2} &=& \nu^2 \gamma \left(\dfrac{\alpha}{\xi_f^3}+\dfrac{\beta(1-\alpha)}{\xi_s^3}\right) \xi_s,\quad&\xi_s(0)=1,&\quad&\dfrac{\textrm{d}\xi_s}{\textrm{d}t}(0)=0&
\end{matrix}
\right . \label{system_to_solve2}
\end{eqnarray}
being 
\begin{equation}
\nu=\left[ \dfrac{4 \pi q_f^2 (n_{f,0}+n_{s,0})}{3 m_f}\right]^{1/2}, \quad
\alpha=\dfrac{n_{f,0}}{n_{f,0}+n_{s,0}}, \quad \beta=\dfrac{q_s}{q_f}.
\end{equation}
The quantity $\nu$ represents a characteristic frequency for the fast ion expansion, while $\alpha \in (0,1)$ is the fraction of fast ions in the cluster. 

\section{Expansion in the outer zone and shock shells}\label{sec:Expansion in the outer zone and shock shells}
The system of equations \eqref{system_to_solve2} describes the dynamics of slow ions during the whole explosion and the motion of fast ions as long as they are inside $S_s$. A fast ion, initially at $r=r_0\leq R$, reaches the border of the inner region when $r_0 \xi_f\left(t\right)=R \xi_s\left(t\right)$. The time the ion crosses the sphere $S_s$, indicated in the following as $\tau\left(r_0\right)$, is obtained by solving the equation
\begin{equation}
\frac{\xi_s(\tau)}{\xi_f(\tau)}=\frac{r_0}{R} .
\label{xi}
\end{equation} 
Considering a fast ion initially at $r_0$, the electric field for $r \geqslant R_s$ can be written as $Q\left(r,t\right)/r^2$, where $Q\left(r,t\right)$ is the charge in the sphere of radius $r$ at time $t$; as long as overtakings between fast ions do not occur, $Q\left(r,t\right)$ is the sum of the total charge of slow ions and the charge of the fast ions initially inside a sphere of radius $r_0$:
\begin{equation}
Q\left(r,t\right)=Q(r_0,t)=\frac{4\pi}{3}\left(q_s n_{s,0} R^3+q_f n_{f,0} r_0^3 \right) . \label{charge_out_S2}
\end{equation}
Consequently, for $t>\tau(r_0)$, the equation of motion for a fast ion outside $S_s$  is
\begin{equation}
m_f\frac{\textrm{d}^2 r}{\textrm{d} t^2} = -\frac {\partial}{\partial r} \left( \frac{q_f Q(r_0)}{r} \right), 
\quad
r(\tau) = \xi_f(\tau) r_0,
\quad
\frac{\textrm{d} r}{\textrm{d} t}(\tau) =\frac{\textrm{d} \xi_f}{\textrm{d} t}(\tau) r_0 .
\label{motion_1_out}
\end{equation}
Equation \eqref{motion_1_out} can be integrated in order to obtain the asymptotic kinetic energy of the fast ions, $\epsilon_{\infty}$, as
\begin{equation}
\epsilon_{\infty}=\frac{1}{2}m_f \left \{ \frac{\textrm{d} \xi_f}{\textrm{d} t}\left(\tau\left(r_0\right)\right) r_0   \right\}^2 +\frac{q_f Q(r_0)}{\xi_f(\tau) r_0} .
\label{energy_as}
\end{equation}
The energy spectrum of the fast ions, $\rho_\epsilon$, can be written as
\begin{equation}
\rho_\epsilon=\dfrac{1}{N_f}\dfrac{\textrm{d} N_f}{\textrm{d} \epsilon_{\infty}}
\end{equation} 
where $N_f$ is the total number of fast ions, while $\textrm{d} N_f$ represents the number of ions having asymptotic energy in the interval $\textrm{d}\epsilon_{\infty}$. Making use of Eqs. \eqref{xi} and \eqref{energy_as}, the spectrum can be expressed parametrically as a function of $\tau$, as:
\begin{equation}
\rho_\epsilon(\epsilon_{\infty}(\tau))=  \dfrac{4 \pi r_0^2(\tau) n_{f,0}}{N_f} \text{  }\dfrac{ \textrm{d}r_0/\textrm{d}\tau}{\textrm{d}\epsilon_{\infty}/\textrm{d}\tau}.
\end{equation}

\begin{figure}
 \begin{minipage}[b]{0.45\textwidth}
   \centering
   \includegraphics[width=\textwidth]{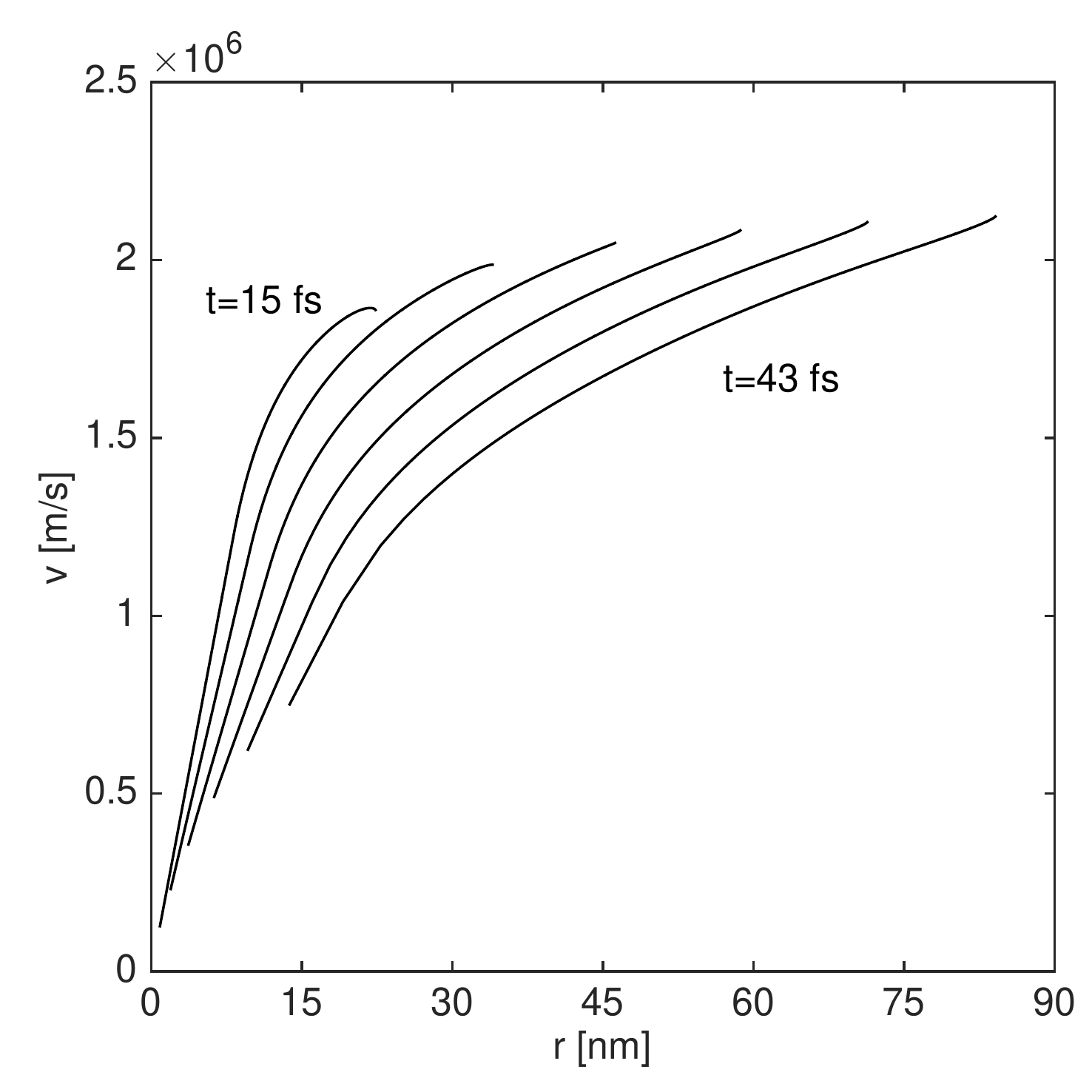}
   \caption{Time evolution of the phase space distribution of fast ions in a mixture $C^+H^+$ with $\alpha=0.4$, for $t$ in the range 15-43 fs. A cluster with ion density $n_{f,0}+n_{s,0}$=10$^{23}$ cm$^{-3}$ and $R$ = 6.5 nm has been considered.}  
   \label{noshock}
 \end{minipage}
 \quad
 \ \hspace{1mm}\
 \begin{minipage}[b]{0.45\textwidth}
   \centering
   \includegraphics[width=\textwidth]{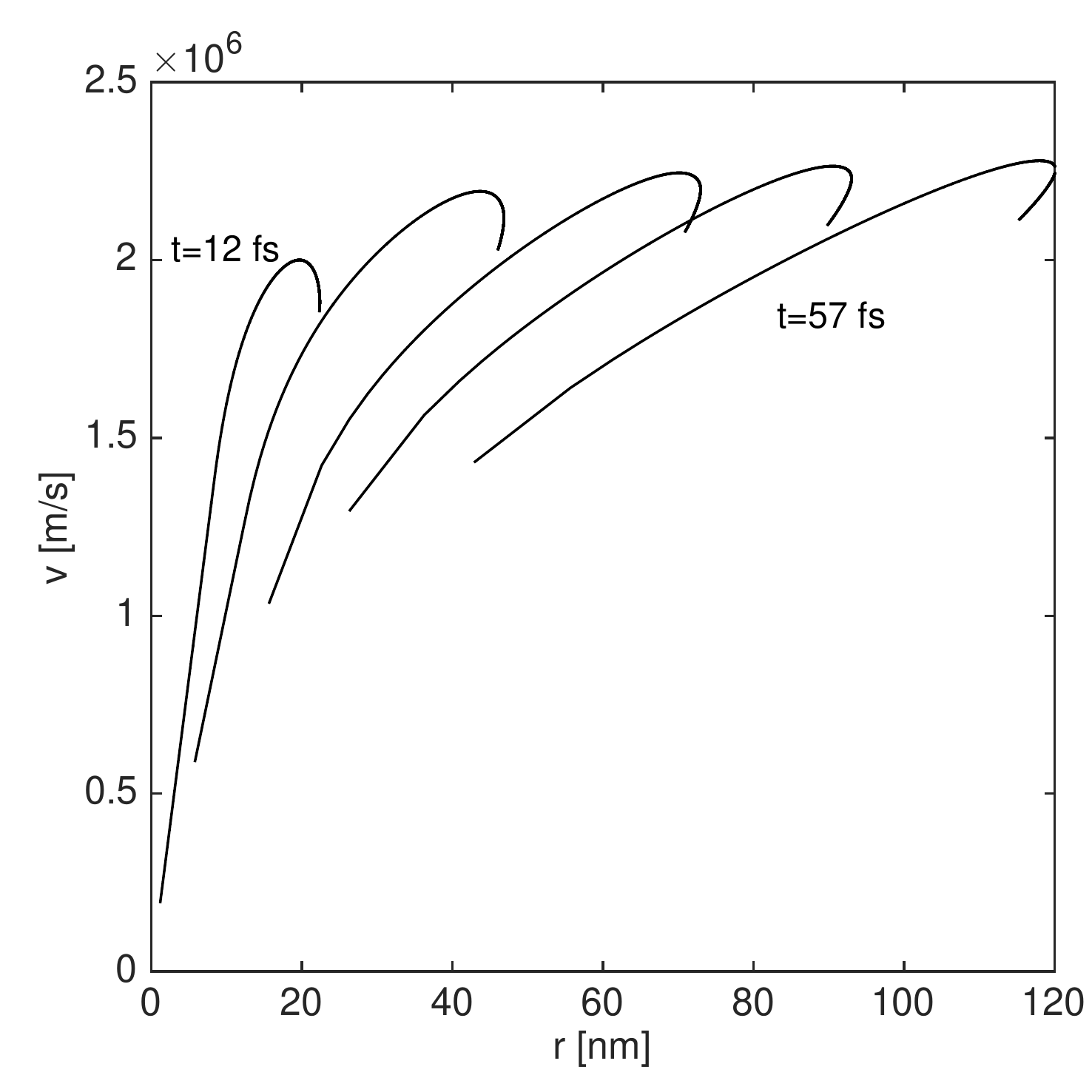}
   \caption{Time evolution of the phase space distribution of fast ions in a mixture $C^+H^+$ with $\alpha=0.2$, for $t$ in the range 12-57 fs. A cluster with ion density $n_{f,0}+n_{s,0}$=10$^{23}$ cm$^{-3}$ and $R$ = 6.5 nm has been considered.}
   \label{shock}
 \end{minipage}
\end{figure}

As pointed out in a previous work \citep{Li-Liu-2007}, shock shells arise when the fraction of fast ions, $\alpha$, is smaller than a critical value, $\alpha_{crit}$. Typical phase-space evolutions of the fast ions are showed in Fig. \ref{noshock} for $\alpha>\alpha_{crit}$ and in Fig. \ref{shock} for $\alpha<\alpha_{crit}$. In particular, the curves in Fig. \ref{shock} show the typical wave-breaking behaviour that indicates the presence of a shock \citep{Kaplan-PRL-2003}.

In fact, there is a simple way to discriminate Coulomb explosions without shocks from the ones in which this phenomenon appears without requiring a full numerical simulation of the explosion. Starting from Eq. \eqref{energy_as}, which, in principle, is valid only when shocks are absent, the derivative of ${\epsilon_{\infty}}$ with respect to $r_0$ is calculated. If the condition
\begin{equation}
\dfrac{\textrm{d}\epsilon_{\infty}}{\textrm{d}r_0} > 0
\label{maggiore}
\end{equation}
is verified for every $r_0 \in [0,R]$, inner ions take a slower velocity respect to the ones in the outer zones and cannot reach them.
In this case, no overtaking is present. 
Instead, if ${\textrm{d}\epsilon_{\infty}}/{\textrm{d}r_0}$ is negative for some $r_0$, in principle there will be overtakings. Of course, Eq. \eqref{energy_as} is no longer valid in this case and it cannot be used to calculate the energy spectrum of fast ions, whose dynamics can be solved only numerically. Using this criterion, the existence of a limit value for $\alpha$ can be readily verified numerically. In general, if $\alpha \in  (\alpha_{crit},1)$ the derivative ${\textrm{d}\epsilon_{\infty}}/{\textrm{d}r_0}$ remains positive and no shock is formed; for $\alpha \in (0,\alpha_{crit})$, ${\textrm{d}\epsilon_{\infty}}/{\textrm{d}r_0}$ changes its sign, meaning that the fast particles overtake each other, giving rise to shock shells. The calculations show that $\alpha_{crit}$ depends only on the ratio between the ions charges and it is independent of their masses \citep{Li-Liu-2007}. In particular, $\alpha_{crit}=1/3$ for mixtures $HD$, $HT$, $DT$ and $HC^+$.

\section{Theoretical derivation of $\alpha_{crit}$}
Starting with the case of no shock, by indicating with $Q_s$ and $Q_f$ the total charge of slow and fast ions, the dynamics of a fast ion initially at $r=r_0$ is governed by the Hamiltonian
\begin{equation}
\mathscr{H}(r, p_r; r_0)=\dfrac{p_r^2}{2m_f}+\dfrac{q_f Q_f}{r}\left(\frac{r_0}{R}\right)^3+q_f \Phi_s(r,R_s(t)) ,
\label{rf_minf}
\end{equation}
where $\Phi_s(r,\rho)$ is the electrostatic potential in $r$ due to fixed ions uniformly distributed in a sphere of radius $\rho$:
\begin{eqnarray}
\Phi_s(r,\rho)=\left \{ \begin{matrix}
\dfrac{3 Q_s}{2 \rho}-\dfrac{Q_s r^2}{2\rho^3} &\textrm{for} & r \leqslant \rho, \\ & & \\
\dfrac{Q_s}{r} &\textrm{for} & r \geqslant \rho.
\end{matrix} \right .
\label{psi_s}
\end{eqnarray}
In the limit situation $m_s\rightarrow+\infty$, $R_s(t)$ is constant and equal to $R$. Consequently, as $\mathscr{H}$ does not depend explicitly on time, it is a constant of motion, and the asymptotic energy $\epsilon_{\infty}$ expressed as a function of $r_0$ can be written immediately as  
\begin{equation}
\epsilon_{\infty} (r_0)=q_f\left[\dfrac{Q_f-Q_s/2}{R^3} r_0^2 + \dfrac{3 Q_s}{2 R} \right] ,
\label{energy_infinity}
\end{equation}
from which the energy spectrum can be readily calculated \citep{Li-Liu-2007, Murakami-PoP-2009}.
\begin{figure}
\centering
\includegraphics[width=0.8\linewidth]{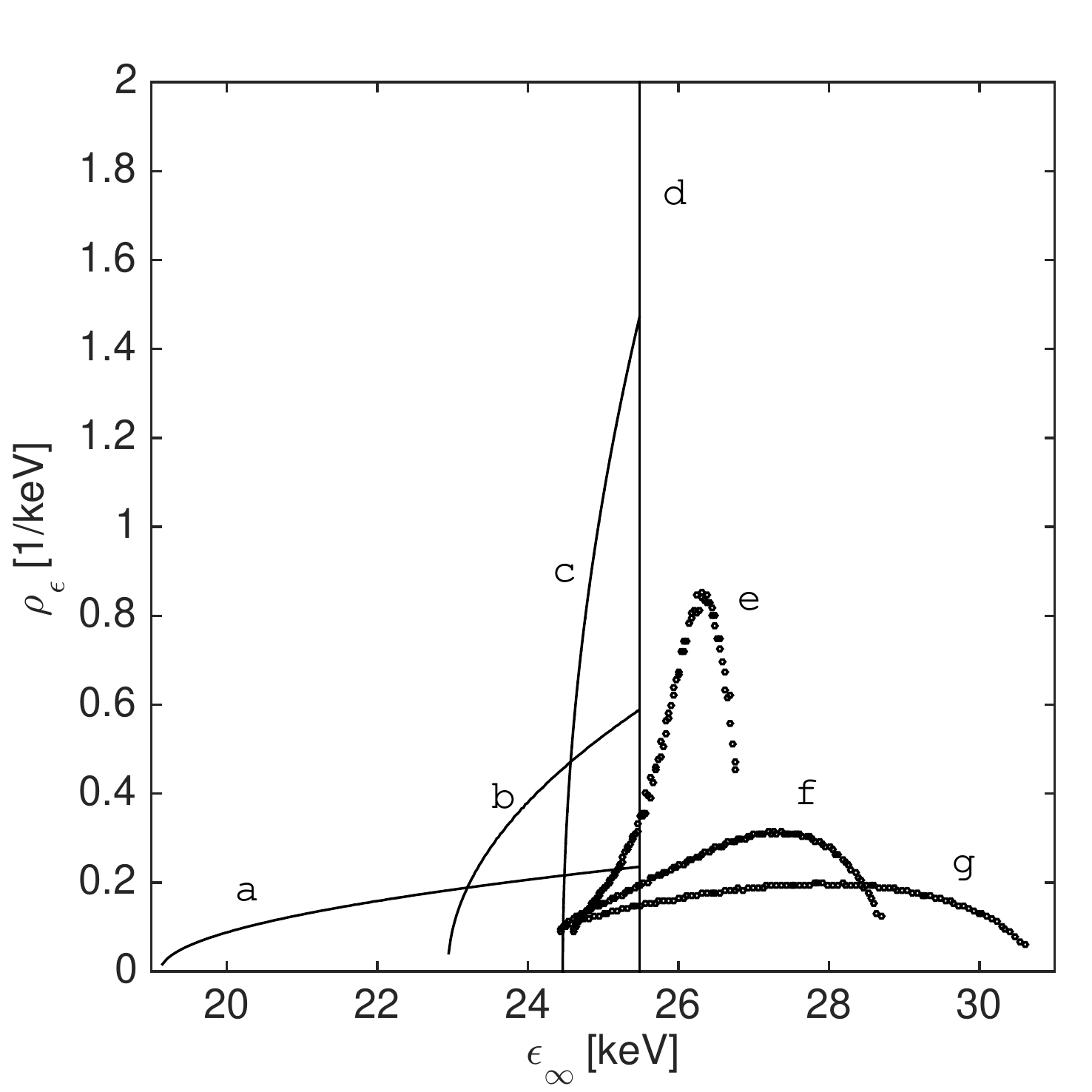}
\caption{Energy spectra of a mixture with slow ions at rest ($m_s\rightarrow+\infty$) and Hydrogen ions, for different value of $\alpha$ in the range $[0.2, 0.5]$. The same initial density and radius as in Fig. \ref{noshock} have been considered. Curves a, b and c represent mixtures with $\alpha>\alpha_{crit}$; in absence of shocks the numerical spectra correspond to the analytic curve. The limit case of $\alpha=\alpha_{crit}$ is shown in curve d, where the energy spectra is singular. Curves e, f and g display situations with $\alpha>\alpha_{crit}$, for which the spectra have been computed by means of the shell method.}\label{spectra}
\end{figure}

Two situations can occur, depending on the sign of $Q_f-Q_s/2$. If $Q_f>Q_s/2$ the ion velocity is an increasing function of $r_0$, so overtakings do not occur, and the analytic spectrum is correct, as shown in Fig. \ref{spectra}, curves a, b, c. When $Q_f<Q_s/2$, according to Eq. \eqref{energy_infinity}, $\epsilon_{\infty}$ is a decreasing function of $r_0$, but this implies that the ions closer to the center of the cluster are bound to overtake the ions close to the border of the sphere. In other words, the hypothesis of no overtakings, and consequently Eq. (\ref{energy_infinity}), are no longer valid. In Fig. \ref{spectra}, curves e, f, g, the energy spectra have been calculated correctly by using the shell method. The case $Q_f=Q_s/2$ (Fig. \ref{spectra}, curve d) can be considered as a limit for $Q_f-Q_s/2\rightarrow0^+$ and, consequently, Eq. \eqref{energy_infinity} can still be used.
In the general case, where the mass of slow ions is finite, in the Hamiltonian \eqref{rf_minf}  the potential due to slow ions, $\Phi_s$, is a function of time, as it depends on the frontline $R_s(t)$,
and  $\mathscr{H}$ is no longer a constant of motion:
\begin{equation}
\dfrac{\textrm{d}\mathscr{H}}{\textrm{d}t}=\dfrac{\partial \mathscr{H}}{\partial t}=q_f \dfrac{\partial \Phi_s}{\partial R_s} \dfrac{\textrm{d}R_s}{\textrm{d}t}.
\label{hamiltonian}
\end{equation}
In principle, the value of the asymptotic energy can be obtained by integrating Eq. (\ref{hamiltonian}) in time: 
\begin{equation}
\epsilon_{\infty}=\mathscr{H}(t \rightarrow +\infty)=\mathscr{H}(t=0)+q_f \int_{0}^{\tau(r_0)} \dfrac{\partial \Phi_s}{\partial R_s}(r(t),R_s(t))\dfrac{\textrm{d} R_s}{\textrm{d} t} \textrm{d}t
\label{integral}
\end{equation}
where the upper integration limit can be set equal to $\tau(r_0)$, as for $t>\tau$ the ion is outside $S_s(t)$ and the term $\partial \Phi_s / \partial R_s$ vanishes. Noticing that $\mathscr{H}(t=0)$ is equal to the asymptotic energy for fixed slow ions, Eq. \eqref{rf_minf}, and expressing $r(t)$ and $R_s(t)$ as functions of $\xi_f$ and $\xi_s$, one finally obtains 
\begin{equation}
\epsilon_{\infty}=q_f\left[\dfrac{Q_f-Q_s/2}{R^3} r_0^2 + \dfrac{3 Q_s}{2 R} \right]-\dfrac{3 Q_s q_f}{2 R} I(r_0)
\label{energy_finite}
\end{equation}
where $I(r_0)$ is defined as
\begin{equation}
I(r_0)=\int_{0}^{\tau(r_0)} \left[1-\left( \dfrac{r_0 \xi_f(t)}{R \xi_s(t)}\right)^2 \right] \dfrac{\xi_s'(t)}{\xi_s^2(t)} \textrm{d}t.
\label{I0}
\end{equation}
The quantity $I$ is always non negative, and, in fact, the motion of slow ions takes some energy away from the fast ones.\\

The condition $\textrm{d} \epsilon_\infty/\textrm{d}r_0>0$ for $r_0 \in [0,R]$ is now investigated. Starting from Eq. \eqref{I0}, one obtains: 
\begin{equation}
\dfrac{\textrm{d} I}{\textrm{d} r_0}=\dfrac{\xi_s'(\tau)}{\xi_s(\tau)^2}\left( 1- \dfrac{r_0^2\xi_f(\tau)^2}{R^2\xi_s(\tau)^2}\right) \dfrac{\textrm{d}\tau}{\textrm{d}r_0}-\int_{0}^{\tau(r_0)} \dfrac{2r_0 \xi_f^2 \xi_s'}{R^2\xi_s^4} \textrm{d}t .
\end{equation}
From this formula, considering that $\textrm{d}\tau/\textrm{d}r_0<0$, one has 
\begin{equation}
\begin{matrix}
\dfrac{\textrm{d}I}{\textrm{d}r_0} (r_0) \leqslant 0 \quad  \forall r_0 \in [0,R],\quad
\dfrac{\textrm{d}I}{\textrm{d}r_0} (R ) = 0.
\end{matrix}
\end{equation}
Being
\begin{equation}
\dfrac{d \epsilon_{\infty}}{d r_0}=2 q_f \dfrac{Q_f-Q_s/2}{R^3} r_0 -\dfrac{3 Q_s q_f}{2 R} \dfrac{\textrm{d}I}{\textrm{d}r_0} ,
\label{deriv_energy_finite}
\end{equation}
if $Q_f>Q_s/2$ the derivative $\textrm{d}\epsilon_\infty/\textrm{d}r_0$ is always non negative, as it is the sum of two non-negative quantities. Instead, if $Q_f<Q_s/2$ a sum of a negative term and a non negative term is present in \eqref{deriv_energy_finite}, and in principle nothing can be said about the sign of  $\textrm{d}\epsilon_\infty/\textrm{d}r_0$ for a generic $r_0$. However, $\textrm{d}I/\textrm{d}r_0$ vanishes for $r_0=R$, and therefore $\textrm{d}\epsilon_\infty/\textrm{d}r_0(R)$ must be negative; consequently, in this case there are overtakings.\\

 In summary, the condition  $Q_f=Q_s/2$ is valid in any case and can be used in order to discriminate between explosions with and without shocks. This is the rigorous proof of a property of the Coulomb explosions of mixtures that was found numerically in a previous work \citep{Li-Liu-2007}. The condition $Q_f=Q_s/2$  provides the critical value of $\alpha$:
\begin{equation}
\alpha_{crit}=\dfrac{\beta}{2+\beta} \textrm{ },
\end{equation}
which depends only on the charge ratio.

\section{Energy spectrum of the fast ions}
\begin{figure}
 \begin{minipage}[b]{0.45\textwidth}
   \centering
   \includegraphics[width=\textwidth]{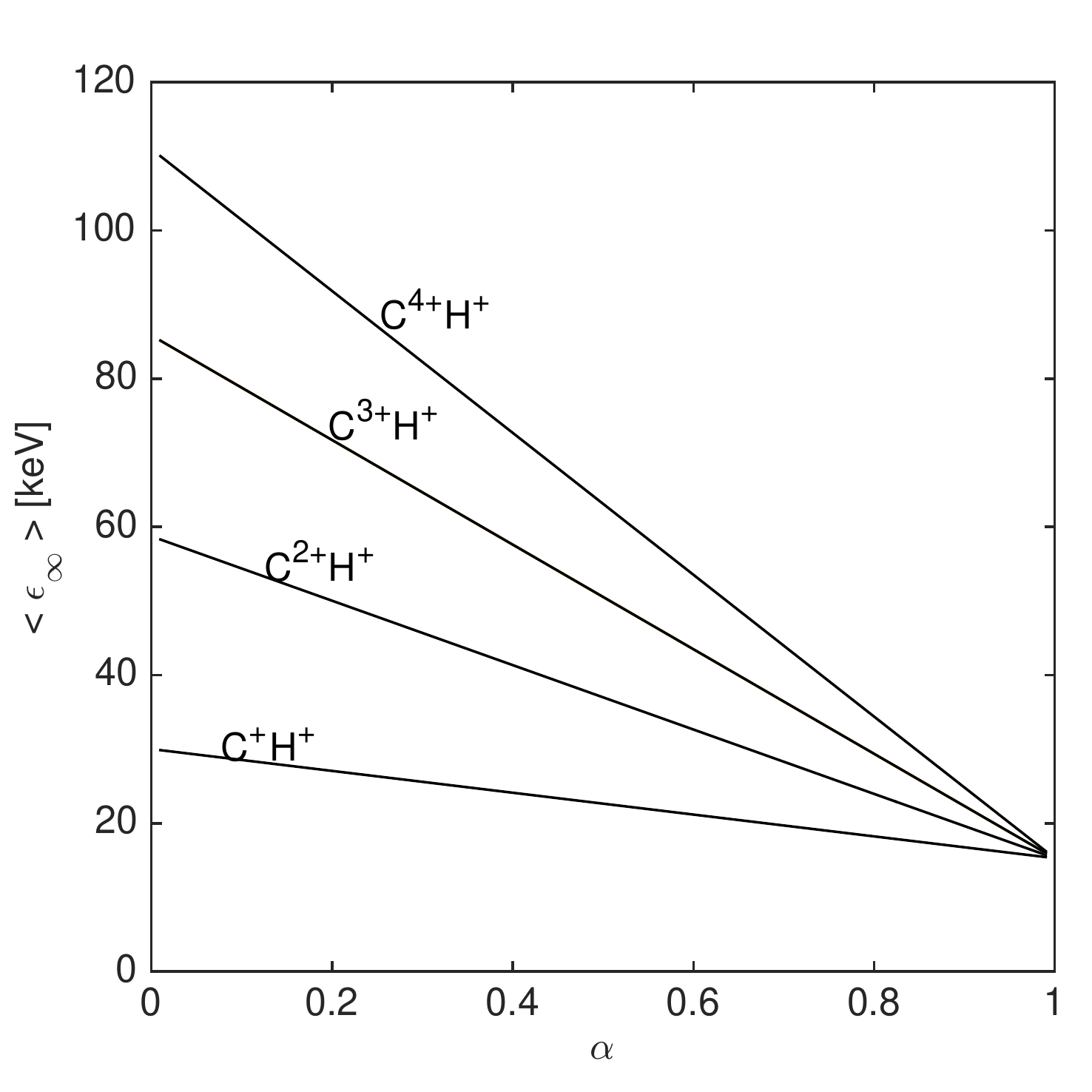}
   \caption{Mean value of the asymptotic energy for different mixtures Carbon-Hydrogen, as a function of the fraction of fast ions.}
   \label{fig:Emedia_carbonio}
 \end{minipage}
 \ \hspace{1mm} \hspace{1mm} \
 \begin{minipage}[b]{0.45\textwidth}
  \centering
   \includegraphics[width=\textwidth]{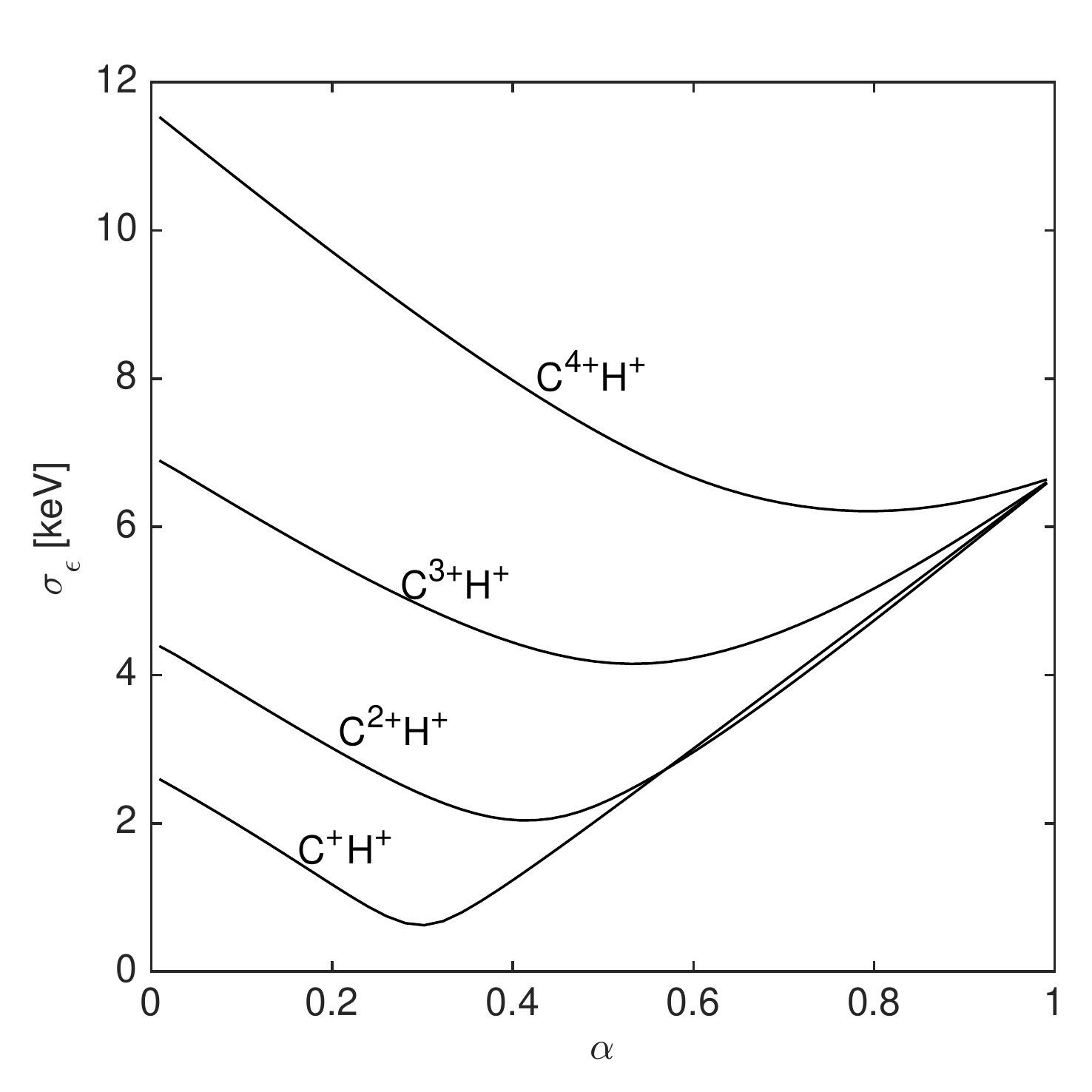}
   \caption{Standard deviation of the energy spectra as a function of the fraction of fast ions, for different mixtures Carbon-Hydrogen.}
   \label{fig:std_dev_carbonio}
 \end{minipage}
\end{figure}

\begin{figure}
 \begin{minipage}[b]{0.45\textwidth}
   \includegraphics[width=\textwidth]{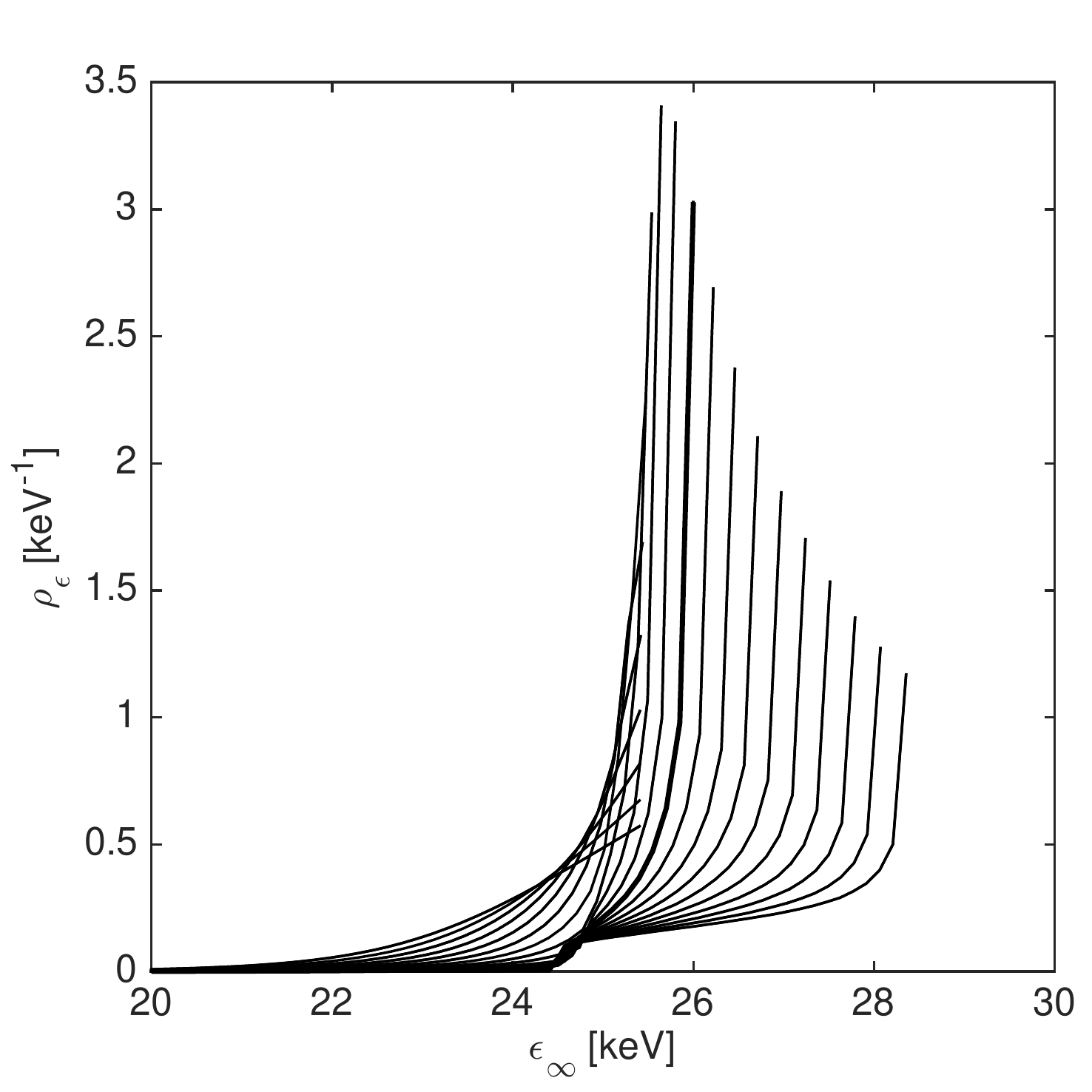}
   \caption{Energy spectra of a mixture $C^+H^+$ with different values of $\alpha$ in the range $[0.2, 0.4]$. The thick curve indicates the distribution with the minimum value of $\sigma_\epsilon$.}
   \label{fig:C1}
 \end{minipage}
 \ \hspace{1mm} \hspace{1mm} \
 \begin{minipage}[b]{0.45\textwidth}
   \includegraphics[width=\textwidth]{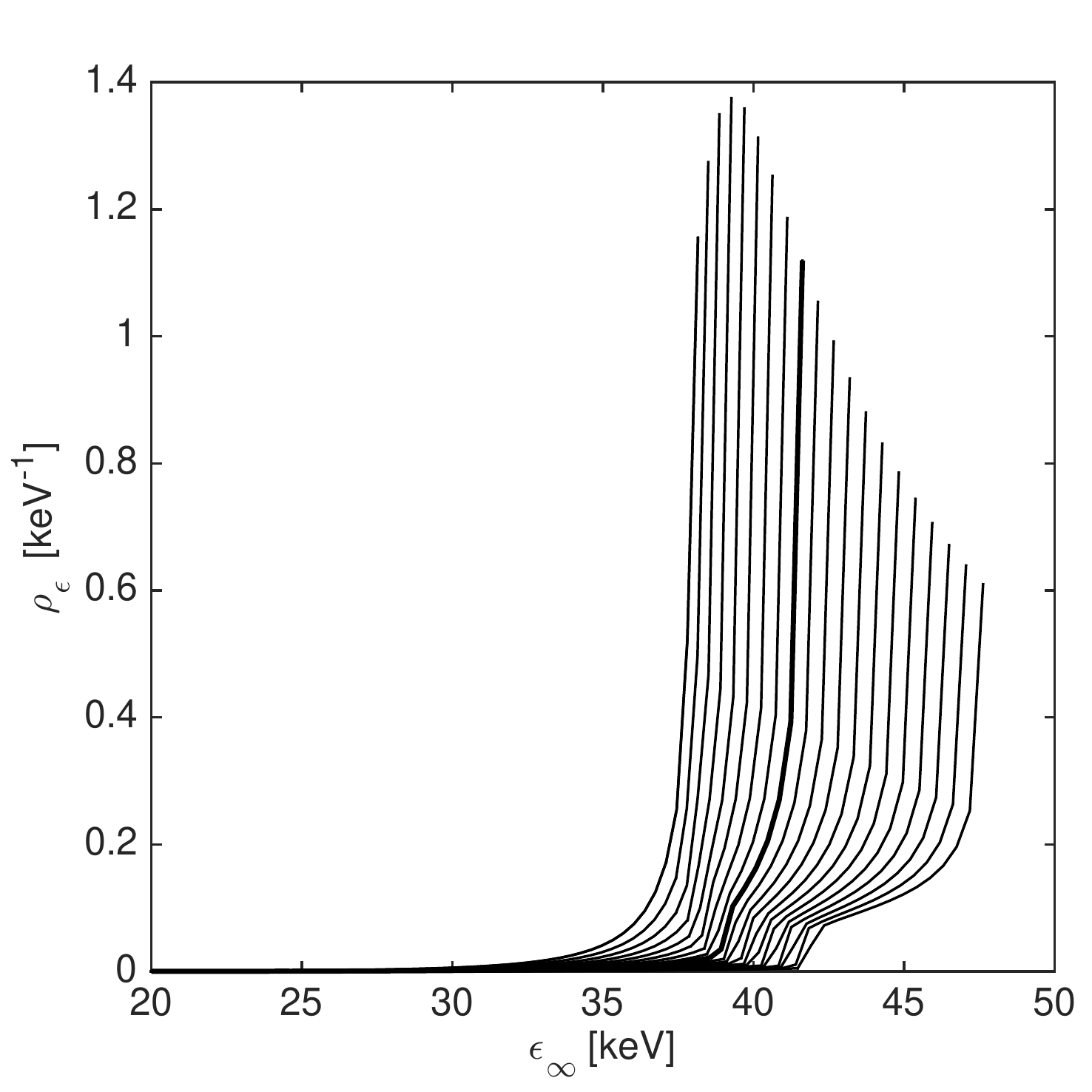}
   \caption{Energy spectra of a mixture $C^{2+}H^+$ with different values of $\alpha$ in the range $[0.3, 0.5]$. The thick curve indicates the distribution with the minimum value of $\sigma_\epsilon$.}
   \label{C2}
 \end{minipage}
\end{figure}

\begin{figure}
 \begin{minipage}[b]{0.45\textwidth}
   \includegraphics[width=\textwidth]{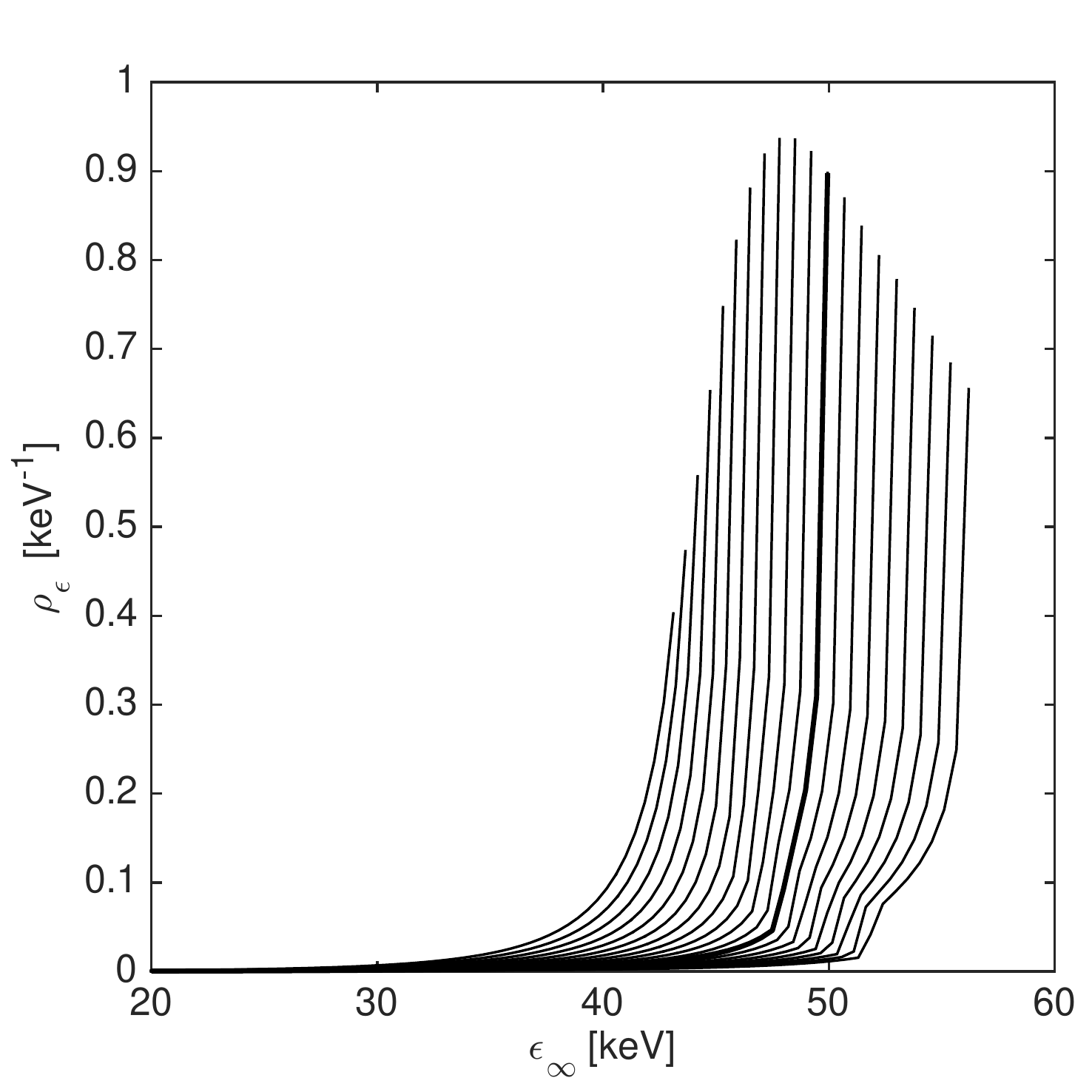}
   \caption{Energy spectra of a mixture $C^{3+}H^+$ with different values of $\alpha$ in the range $[0.45, 0.65]$. The thick curve indicates the distribution with the minimum value of $\sigma_\epsilon$.}
   \label{C3}
 \end{minipage}
 \ \hspace{1mm} \hspace{1mm} \
 \begin{minipage}[b]{0.45\textwidth}
   \includegraphics[width=\textwidth]{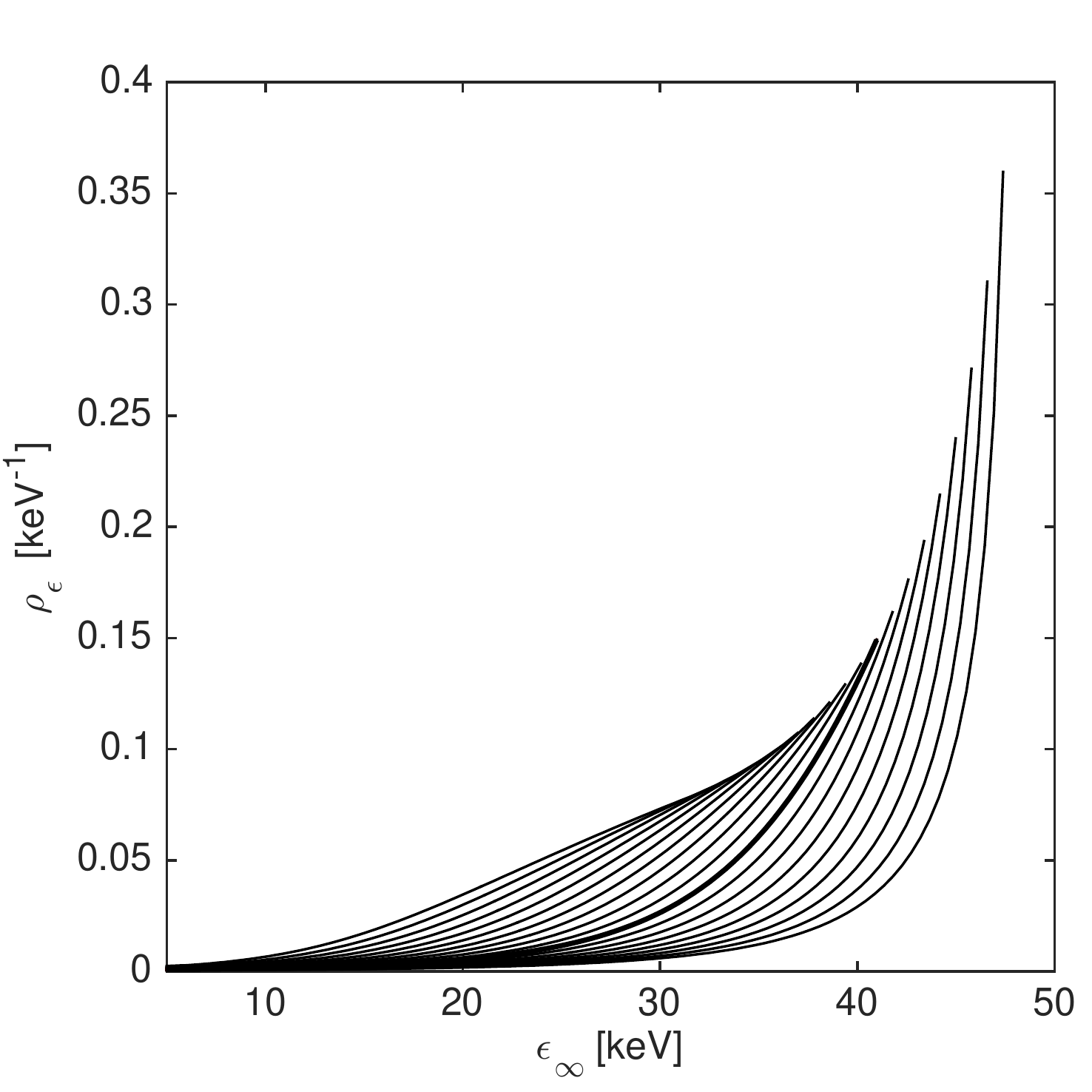}
   \caption{Energy spectra of a mixture $C^{4+}H^+$ with different values of $\alpha$ in the range $[0.7, 0.9]$. The thick curve indicates the distribution with the minimum value of $\sigma_\epsilon$.}
   \label{fig:C4}
 \end{minipage}
\end{figure}

The analysis performed for the simplified situation of $m_s \rightarrow +\infty$ suggest that fast ion spectra could be particularly narrow when $\alpha$ approaches $\alpha_{crit}$. To confirm that, a full set of calculations has been carried out for the mixtures $C^+H^+$, $C^{2+}H^+$, $C^{3+}H^+$ and $C^{4+}H^+$. Clusters with ion density $n_{f,0}+n_{s,0}$=10$^{23}$ cm$^{-3}$ and $R$ = 6.5 nm (as in Figs. \ref{noshock}-\ref{spectra}) have been considered. In Figs. \ref{fig:Emedia_carbonio} and \ref{fig:std_dev_carbonio} the mean value of the energy and the energy spread have been calculated for each type of mixture for $\alpha \in (0,1)$. Moreover, in Figs. \ref{fig:C1}-\ref{fig:C4} the energy spectra are reported for each mixture for $\alpha$ close to $\alpha_{crit}$. All the calculations have been performed with the shell method.\\

The plots of the mean ion energy, $\left<\epsilon_\infty\right>=\int_{0}^{+\infty} \epsilon \rho_{\epsilon}(\epsilon)\textrm{d}\epsilon $, as function of $\alpha$, show a linear behaviour. In fact, the total kinetic energy of the fast ions, $\mathscr{E}_f$, can be written as the difference between the potential energy at $t=0$ ad the total kinetic energy of the slow ions
\begin{equation}
\mathscr{E}_f=\mathcal{U}(t=0)-\mathscr{E}_s=\dfrac{3(Q_s+Q_f)^2}{2R}-\dfrac{3}{5}N_s \dfrac{m_s}{2}[R \xi_s'(+\infty)]^2
\end{equation}
and, therefore, it does not depend on the possible presence of shocks. 

From the numerical solution of system \eqref{system_to_solve}, it has been found out that $[\xi_s'(+\infty)]^2$ depends linearly on $\alpha$, and using this result the average kinetic energy $\left<\epsilon_\infty\right>=\mathscr{E}_f/N_f$ becomes a linear function of $\alpha$.



The standard deviation of the energy of fast ions, $\sigma_\epsilon=[\int_{0}^{+\infty}( \epsilon-\left<\epsilon_\infty\right>)^2 \rho_{\epsilon}\textrm{d}\epsilon]^{1/2} $, presents a minimum for a value of $\alpha$ which is close to $\alpha_{crit}$ when $(m_s/q_s)/(m_f/q_f)\gg1$. In fact, for $\alpha \simeq \alpha_{crit}$ the spectrum has a large plateau and a sharp peak. Probably, the plateau gives a non negligible contribution to $\sigma_\epsilon$, while for practical applications the presence of the peak and its energy spread are more important.


%

\section{Summary}
In the paper, the dynamics of particles during the Coulomb explosion of ion mixtures has been investigated, with particular interest to the energy spectrum of the component with the larger charge-to-mass ratio. Exploiting the hypothesis of no overtakings among ions of the same species and the new approach based on the linear behavior for the electric field inside the inner sphere, $S_s$, containing both slow and fast ions, analytical formulas describing the acceleration both inside  and outside $S_s$ (where only fast ions are present), have been derived together with the energy spectrum. For the first time (at least to the Authors' knowledge), a simple way to deduce if overtakings between fast ions will take place during the expansion is presented and discussed, confirming the existence, empirically deduced in a previous work by using numerical simulations \citep{Li-Liu-2007}, of a limit value for the mixture composition, $\alpha_{crit}$, above which no shocks are forming. The analysis here proposed is based on the calculation of the derivative of $\epsilon_{\infty}$ with respect to the initial ion position $r_0$: when this quantity is negative for some $r_0$, the hypothesis of no overtaking is no longer valid. Moreover, making use of the Hamiltonian of fast particles and starting  with the simple case where the mass of the slow species is considered infinite, a rigorous proof of the existence of $\alpha_{crit}$, which is valid for every value of $m_f$ and $m_s$, has been obtained.  In the limit situation of $m_s\rightarrow+\infty$, when $\alpha>\alpha_{crit}$  approaches the critical value, the energy spectrum becomes narrower and narrower, meaning that the ions tend to acquire very similar velocities, and for $\alpha=\alpha_{crit}$ the spectrum becomes monoenergetic. On the other hand, the condition $\alpha=\alpha_{crit}$ is the limit situation for the rising of shock shells. As most of these considerations are valid also for the general case in which $m_s$ is finite, one can conclude that the rising of shock shells is induced by a very narrow energy spectrum, and not vice versa.

In conclusion, conditions to obtain nearly monoenergetic ions from Coulomb explosions of heteronuclear clusters have been individuated and supported by a rigorous theoretical analysis. The analysis has been carried out supposing that the electrons are completely expelled from the cluster; also in order to explain recent experimental results \citep{Hohenberger-PRL-2005, Iwan-PRA-2012}, future investigation should take into proper account also the electron dynamics and the laser-cluster interaction in order to have a clearer picture of the phenomenon. \\

This work was partially supported by the European Research Council (ERC-2010-AdG Grant No. 267841).


\end{document}